\documentclass[aps,prl,twocolumn,longbibliography,10p]{revtex4-2}
\usepackage[utf8]{inputenc}

\usepackage{times}
\usepackage{units}
\usepackage{physics}
\usepackage{graphicx}
\usepackage[normalem]{ulem}

\usepackage[vietnamese,english]{babel}

\usepackage{hyperref}
\hypersetup{
colorlinks=true, 
linkcolor=[RGB]{45,48,146}, 
citecolor=[RGB]{45,48,146}, 
filecolor=[RGB]{45,48,146}, 
urlcolor=[RGB]{45,48,146}}  

\newcommand{\Fmf}[2]{\left| F=#1, m_F=#2 \right\rangle}

\begin{document}

\title{A Quantum Repeater Node Demonstrating Unconditionally Secure Key Distribution}

\author{S. Langenfeld}
\email[To whom correspondence should be addressed. Email: ]{stefan.langenfeld@mpq.mpg.de}
\affiliation{Max-Planck-Institut f\"{u}r Quantenoptik, Hans-Kopfermann-Strasse 1, 85748 Garching, Germany}
\author{P. Thomas}
\affiliation{Max-Planck-Institut f\"{u}r Quantenoptik, Hans-Kopfermann-Strasse 1, 85748 Garching, Germany}
\author{O. Morin}
\affiliation{Max-Planck-Institut f\"{u}r Quantenoptik, Hans-Kopfermann-Strasse 1, 85748 Garching, Germany}
\author{G. Rempe}
\affiliation{Max-Planck-Institut f\"{u}r Quantenoptik, Hans-Kopfermann-Strasse 1, 85748 Garching, Germany}

\begin{abstract}
\noindent Long-distance quantum communication requires quantum repeaters to overcome photon loss in optical fibers. Here we demonstrate a repeater node with two memory atoms in an optical cavity. Both atoms are individually and repeatedly entangled with photons that are distributed until each communication partner has independently received one of them. An atomic Bell-state measurement followed by classical communication serves to establish a key. We demonstrate scaling advantage of the key rate, increase the effective attenuation length by a factor of two, and beat the error-rate threshold of 11\% for unconditionally secure communication, the corner stones for repeater-based quantum networks.
\end{abstract}

\maketitle

\noindent Optical repeaters based on light amplification have been a game changer for the development of modern telecommunication links, enabling fiber-based networks at the global scale. The same is to be expected for their quantum-physical counterparts, namely quantum repeaters. In both cases, the issue is to overcome the propagation loss in long-distance transmission lines. As quantum signals, the qubits, cannot be amplified or copied \cite{Wootters1982}, the classical repeater strategy fails for quantum links. The challenge was resolved by Briegel et al.\ \cite{Briegel1998} who proposed a quantum repeater protocol that can distribute entanglement, the basic resource for quantum networks, between quantum memories \cite{Kimble2008,Wehner2018}.

The main idea behind a repeater \cite{Briegel1998,Duan2001} is to replace the probabilistic transmission through the quantum channel by a heralded preparation of the quantum link followed by deterministic classical communication. Towards this goal, the link is divided into distinct segments connected by repeater nodes. These nodes serve to independently prepare each segment in an entangled state that can be used for communication. The subsequent concatenation of all segments by entanglement swapping then improves the rate-versus-distance scaling for the channel transmission in a fundamental way. Of course, this repeater advantage is in vain if the efficiency is small and the errors are large. Hence, high-fidelity operations are essential for quantum repeaters. Moreover, as the preparation of the segments is achieved probabilistically with a heralded repeat-until-success strategy, synchronization of all segments requires repeater nodes with long-lived qubit memories.

The necessary elements for a scalable quantum repeater such as light-matter entanglement \cite{Duan2010}, qubit memories \cite{Heshami2016}, Bell-state measurements (BSMs) \cite{Pirandola2015}, entanglement swapping \cite{Pan1998} and distillation \cite{Kalb2017} have been investigated individually for different platforms and a plethora of protocols \cite{Yuan2008,Sangouard2011,Pu2021}. However, up to date, there has been no experimental demonstration of the combination of these ingredients into a single repeater protocol, mostly due to technical limitations concerning efficiency and fidelity, incompatibility of different qubit carriers, or irreconcilability of the individual steps of the protocol.

Reaching the goal of a repeater-increased communication distance remains a grand challenge that needs to be addressed step by step. Quantum key distribution (QKD) provides an ideal and application-friendly setting for this approach \cite{Xu2020}, with the practical advantage that end nodes are implemented as classical parties, Alice and Bob, instead of quantum memories. In QKD, two parties establish a secret key that is unconditionally secure against attacks by adversary eavesdroppers provided the quantum bit error rate (QBER), i.e.\ the infidelity of the quantum link, remains below the threshold of $\unit[11]{\%}$ \cite{Shor2000}. A useful corollary is that distillation can be performed classically on the obtained key. Towards surpassing the rate-versus-distance scaling of direct transmission, specific protocols were proposed \cite{Panayi2014,Luong2016,Lucamarini2018}, and various platforms were analyzed \cite{vanLoock2020}, but experimental work is still elusive: One demonstration showed an improved scaling but is fundamentally limited to a single node \cite{Minder2019}, another employs a memory but the required QBER was not achieved and the scalability to multiple nodes remains uncertain \cite{Bhaskar2020}.

Here we combine state-of-the-art quantum-optical techniques to experimentally approach the protocol proposed by Luong et al.\ \cite{Luong2016}. This protocol examines a modular building block that can be concatenated to construct a quantum repeater which, considering only channel losses, is scalable to larger distances \cite{Borregaard2015b,Uphoff2016,vanLoock2020}. We realize the core element, dubbed a quantum repeater node, with two segments that connect to Alice and Bob. It achieves the necessary QBER to distribute unconditionally secure keys with a probability that scales with distance more favorable, i.e.\ proportional to the square root of the direct transmission probability. The implementation relies heavily on the toolbox provided by cavity quantum electrodynamics (QED) \cite{Reiserer2015}. Most importantly, the optical cavity acts as a light-matter quantum interface for the efficient generation of atom-photon entanglement \cite{Wilk2007}, and two individually and repeatedly addressable intracavity atoms serve as two distinct high-fidelity qubit memories \cite{Langenfeld2020}. The BSM again relies on cavity QED and the common cavity mode in which the atoms are localized. A further advantage for future experiments is the demonstrated possibility to perform quantum-logic gates \cite{Welte2018}, e.g., for entanglement purification.

\begin{figure}
\includegraphics[width=1.0\columnwidth]{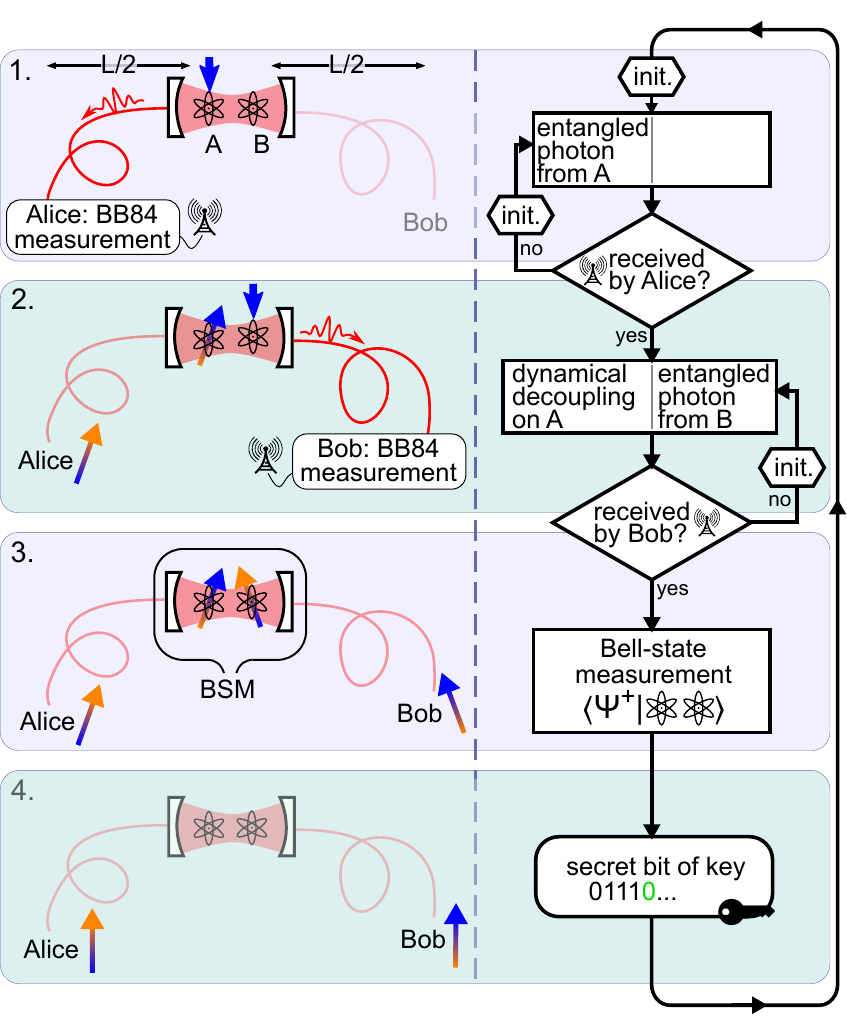}
\caption{Quantum-repeater scheme for distributing a string of secret bits. Two atoms in a cavity serve as a repeater node. Atom A repeatedly creates atom-photon entanglement and sends the photon to Alice until she announces (antenna) the detection of a photon in a BB84 measurement (1), thus creating a correlation between atom A and Alice (indicated by a color-correlated spin pair of same angle). The same for atom B and Bob (2) while the qubit stored in atom A is protected by dynamical decoupling. A local Bell-state measurement (BSM) on the two atoms swaps the two correlations to Alice and Bob (3), who share a secret key after classical post-processing (4).}
\end{figure}

As depicted in Fig.\ 1, the experimental sequence is divided into four parts \cite{Luong2016}: First, atom A is initialized and subsequently transferred into an atom-photon polarization-entangled pair. The photon travels via an optical fiber from the repeater node to one of the communication partners, Alice, where it is detected in a polarization-resolving measurement setup. Following the BB84 protocol \cite{Bennett1984}, Alice randomly chooses between two non-orthogonal detection bases, thereby establishing a correlation between the state of atom A and the measurement result of Alice. In Fig.\ 1 this is highlighted via color-correlated spins of the same angle. If Alice did not detect a photon, e.g., due to fiber transmission losses, this failure is communicated back to the repeater node in order to restart the sequence with initialization and entanglement generation. 

Conversely, a photon detected by Alice heralds a successful transmission and signals the repeater node to continue with step two. This consists of the same sequence, but now applied to atom B and the other communication partner, Bob. While atom B repeatedly tries to connect to Bob, the previously established correlation between Alice and atom A needs to be preserved. The maximum number of entanglement attempts on atom B, $n$, is therefore limited by the coherence time of the memory atom A and potential cross talk between the two atoms. This number can be increased by extending the qubit coherence time, e.g., by applying dynamical decoupling on atom A while trying to connect atom B with Bob. If within $n$ trials no photon was detected, the whole sequence is aborted and restarts with the initialization of atom A.

After successful detection of two photons, one by Alice and one by Bob, the repeater node carries one qubit in each memory, one correlated with Alice and the other with Bob. In this case the third step of the sequence proceeds with the BSM of the two atoms. This swaps the correlation from atom A-Alice and atom B-Bob to Alice-Bob, leaving no trace of the correlation in the repeater node. Finally, in the forth and last step the result of the BSM is publicly announced to Alice and Bob which then share one more bit of raw, i.e.\ not yet secured, key. A secret key can be obtained after classical error correction and privacy amplification \cite{Scarani2009}.

We emphasize that the described protocol can straightforwardly be extended to distribute entanglement between Alice and Bob by replacing the BB84 photon-absorbing end nodes with heralded qubit memories \cite{Chen2008,Brekenfeld2020}. Most importantly for quantum networks, the protocol is scalable to a chain of repeater nodes connecting Alice and Bob, e.g., by simply interfering photons from neighboring nodes on a beamsplitter \cite{Uphoff2016}. This would further amplify the scaling advantage.

\begin{figure}
\includegraphics[width=1.0\columnwidth]{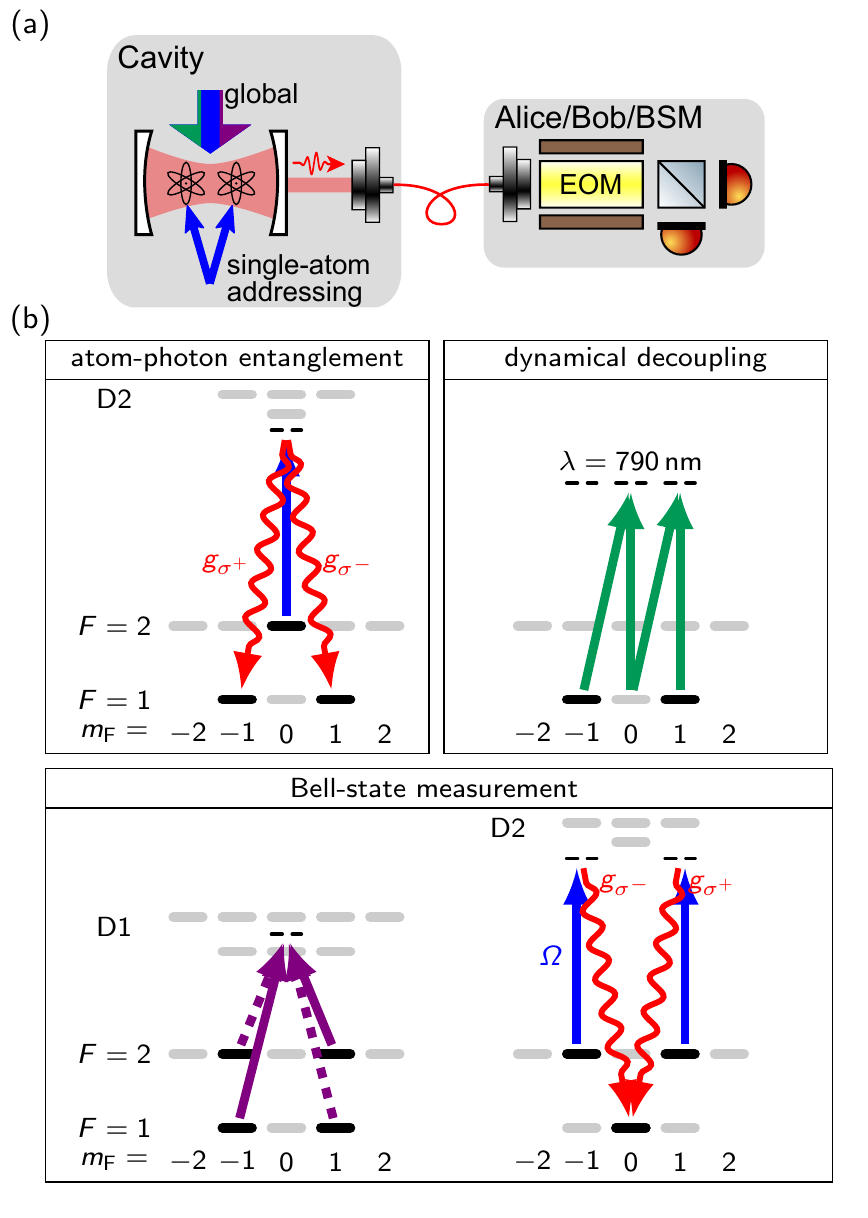}
\caption{Implementation of and toolbox for a quantum-repeater protocol.
(a) Sketch of the experimental setup. Two $^{87}$Rb atoms in a high-finesse cavity serve as matter qubits and can either be addressed individually or globally. A single detection setup is used for Alice, Bob and the photonic BSM with an electro-optical modulator (EOM) for fast polarization/detection basis selection.
(b) Level diagrams and relevant optical fields for the steps described in Fig.\ 1.
}
\end{figure}

The experiment starts by loading two $^{87}$Rb atoms close to the center of the optical cavity where they are trapped in a two-dimensional optical lattice \cite{Supplement}. Both atoms couple about equally to the cavity mode, while classical light fields can be applied globally or individually via an optical addressing system (Fig.\ 2a). In our proof-of-principle experimental demonstration, a single detection setup plays the role of both Alice and Bob. A fast electro-optical modulator (EOM) is used to switch the polarization analysis basis between two of the Pauli eigenbases, i.e.\ $X$ and $Z$. The same detection setup is later on used for the measurement of Bell-states in the $Z$-basis.

The sequence begins by initializing the atom in the ground state $\Fmf{2}{0}$, from which it generates an atom-photon spin-polarization-entangled state via a vacuum-stimulated Raman adiabatic passage (vSTIRAP) \cite{Wilk2007} to $\Fmf{1}{\pm1}$ (Fig.\ 2b). In order to avoid unintended cross talk between the two atoms, we employ a large single-photon detuning of $\unit[-200]{MHz}$ with respect to the excited atomic state $\ket{5^2P_{3/2}, F'=1, m_F=\pm1}$. Moreover, applying the control laser pulse selectively to only the wanted atom avoids cross-illumination between the atoms \cite{Langenfeld2020}. Due to constraints in the atom cooling and trapping \cite{Supplement}, entanglement generation is repeated a maximum of $n$ times. 

In the second step of the protocol (Fig.\ 1), atom B has to be repeatedly re-initialized without affecting the qubit already stored in atom A. This is achieved by atom-selective pumping from $\ket{F=1}$ to $\ket{F=2}$ via the optical addressing system. At the same time, Zeeman pumping to $\ket{m_\mathrm{F}=0}$ is applied globally, i.e.\ without being atom-selective, as the employed transition $\ket{F=2} \leftrightarrow \ket{5^2P_{1/2}, F'=2}$ is $\unit[6.8]{GHz}$ detuned from the qubit-carrying states in $\ket{F=1}$. The initialization is optimized to be fast, to have as little cross talk as possible, but still maintain a high efficiency. 
We achieved a single-trial success probability for zero communication distance, $L=0$, of $p_\mathrm{AB,L=0}=\unit[(22.13\pm0.03)]{\%}$. This approximately matches the combination of the individually obtained values for the atom initialization ($\unit[66]{\%}$), photon generation ($\unit[69]{\%}$), fiber-coupling including optical elements ($\unit[85]{\%}$) and detection efficiency ($\unit[68]{\%}$)  \cite{Supplement}.
The whole atom initialization takes $\unit[8]{\mu s}$, followed by $\unit[2]{\mu s}$ for photon emission and an additional waiting time of up to $\unit[10]{\mu s}$ for receiving the heralding signal from Alice or Bob. Thus, the atom-photon entanglement attempts are repeated every $\unit[20]{\mu s}$.

Qubit coherence time is a very important aspect for any memory-based quantum-repeater architectures \cite{Muralidharan2016}. Here, we use a dynamical decoupling scheme (Fig.\ 2b) to improve the coherence time from below $\unit[1]{ms}$ \cite{Langenfeld2020} to above $\unit[20]{ms}$. This extension is both necessary and sufficient for the protocol implemented here. Details are described in \cite{Supplement}.

After successfully creating correlations in both segments, i.e.\ Alice with atom A and Bob with atom B, a BSM swaps the correlation to Alice and Bob. We perform a linear-optics BSM \cite{Pirandola2015} on photons carrying the qubit information of atoms A and B while the common cavity mode erases the which-way (which-atom) information from the photons. More specifically, in order to drive a vSTIRAP for qubit-readout starting from $\ket{F=2}$, we first map both atomic qubits simultaneously from $\Fmf{1}{\pm1}$ to $\Fmf{2}{\mp1}$ via a two-photon Raman transition (Fig.\ 2b). Afterwards, the vSTIRAP with global control beam generates two photons which are ideally indistinguishable as they originate from atoms in the same cavity mode and are driven by the same control beam. The detection of one photon in state $\ket{\uparrow_Z}$ and the other photon in state $\ket{\downarrow_Z}$ heralds the symmetric Bell-state $\ket{\Psi^+}=(\ket{\uparrow_Z \downarrow_Z}+\ket{\downarrow_Z\uparrow_Z})/\sqrt{2}$ \cite{Supplement} and thus selects those cases where Alice and Bob must have obtained (anti-) correlated results if both had measured in the ($Z$-) $X$-basis. This BSM, which involves the generation and detection of two photons, has an efficiency of $p_\mathrm{BSM}=\unit[(5.07\pm0.03)]{\%}$, including the $\unit[50]{\%}$ detection limitation of linear optics.

\begin{figure}
\includegraphics[width=1\columnwidth]{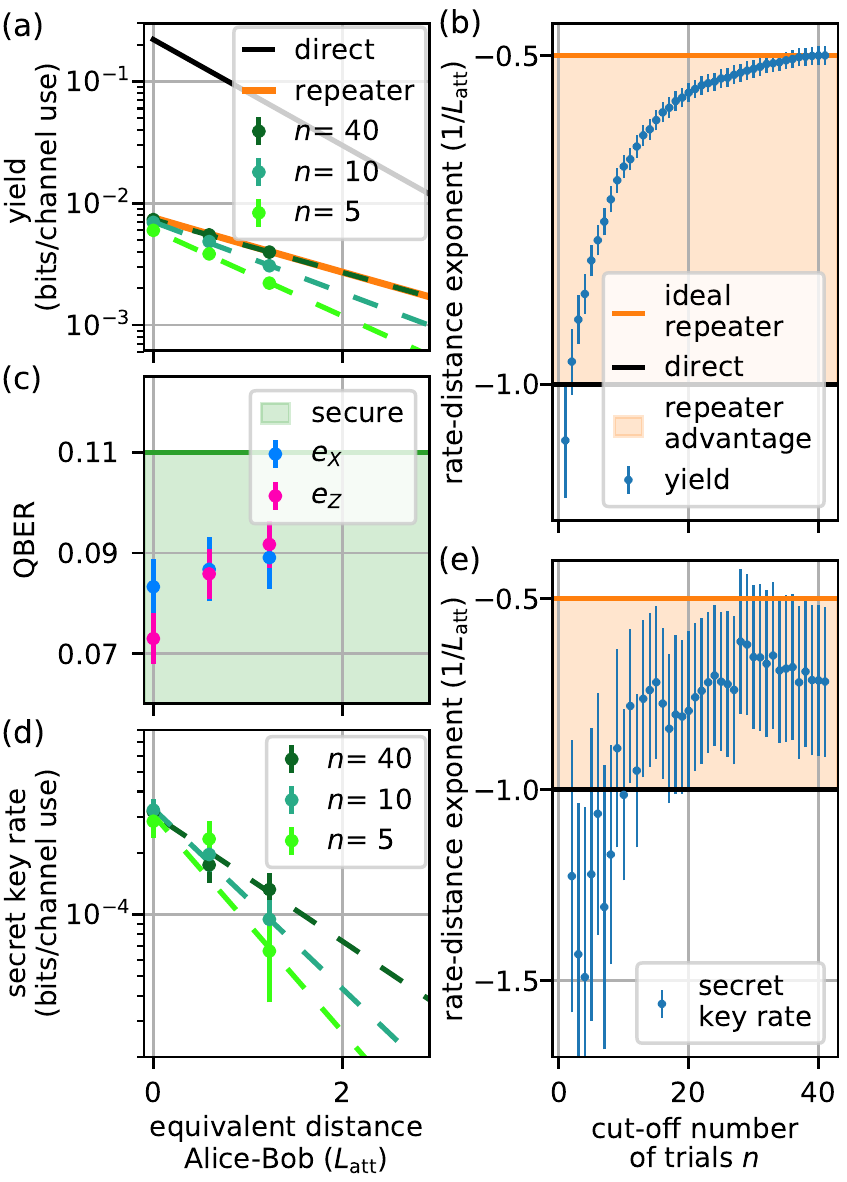}
\caption{Creating an unconditionally secure key with repeater advantage.
(a) Yield of the repeater protocol for different distances and three exemplary cut-offs in the number of trials, $n$. Dashed lines show fits to the data. Theory curves for the repeater case and the direct transmission case serve as comparison.
(b) Slope of the fits given in (a), quantifying the yield-versus-distance scaling as a function of the maximum number of trials, $n$. With increasing $n$, the exponent changes from $-1/L_\mathrm{att}$ (direct transmission) to $-0.5/L_\mathrm{att}$ (repeater advantage).
(c) For both detection bases the quantum bit error rate (QBER) beats the threshold of $\unit[11]{\%}$ for unconditional security.
(d, e) Same as (a, b) for secret key rate. All error bars represent one standard deviation of the statistical uncertainty (details in \cite{Supplement}).
}
\end{figure}

Experimental results are depicted in Fig.\ 3. First, we analyze the yield of the key generation process, i.e.\ how many raw bits are generated per channel use, and also compare it to the achievable rate when using direct transmission with a setup using the same efficiencies for photon generation and detection. In contrast to Ref.\ \cite{Bhaskar2020}, we do not attribute the system inefficiencies to an effective distance. Instead, we consider as equivalent distance only the losses we add on top of our experimental imperfections. This results in a curve starting at $L=0$. Figure 3a shows the achieved yield for different cut-offs in the number of trials $N\leq n$. Following the usual convention \cite{Luong2016}, we evaluate the number of trials as $N=\mathrm{max}(N_\mathrm{A}, N_\mathrm{B})$, i.e.\ the maximum number of trials needed by atom A/Alice and atom B/Bob. For small $n$, i.e.\ without fully utilizing the repeat-until-success strategy, the rate decays very similarly to direct transmission, but with an offset that is given by the unavoidable inefficiency. For $n=40\gg1/p_\mathrm{AB,L}$, i.e.\ effectively without cut-off in the examined range of $L$, the experimental data follow the theoretical expectation with a raw key rate that follows a scaling $\exp(-L/(2L_\mathrm{att}))$ instead of $\exp(-L/L_\mathrm{att})$ for direct transmission. Here, $L_\mathrm{att}$ is the attenuation length of the optical fiber. The scaling advantage is further highlighted in Fig.\ 3b. By increasing the maximum number of trials the data demonstrate a smooth transition from the direct-transmission regime to the memory-assisted regime. Here, the absolute raw key rate is $0.57~\mathrm{bits}/{s}$ \cite{Supplement}.

In order to establish an unconditionally secure key between Alice and Bob, the experiment aims at a QBER below $\unit[11]{\%}$ \cite{Shor2000,Xu2020}. We analyze this by comparing the obtained (anti-) correlation of Alice and Bob with the theoretically expected one. The error rates $e$ are given in Fig.\ 3c for the two chosen BB84 detection bases $X$ and $Z$ which, following the earlier introduced atom-photon entanglement protocol (Fig.\ 2b), correspond to atomic superposition and energy eigenstates, respectively. For all distances, we beat the $\unit[11]{\%}$ threshold by at least three standard deviations. In order to achieve this, we observe that in our BSM the error rate increases with the time separation between the two photon-detection events. Thus only trials with a sufficiently small detection-time difference are used for secret-key generation \cite{Supplement}. This reduces the secret-key rate by about a factor of four. From single-atom single-photon characterization experiments, we estimate that about $\unit[8]{\%}$ of the total QBER is due to infidelities in the atom-photon entanglement generation mechanism governed by our atom-cavity parameters and off-resonant scattering in the employed Raman sequences. Thus, the polarization alignment error and the reduction in visibility of the BSM are negligible in comparison ($\unit[\lesssim 1]{\%}$) using the selection on small detection-time differences explained above.

More fundamentally, the secret key rate is given by the product of yield and secret-key fraction, $r_S$, which is lower bounded by \cite{Shor2000}
\begin{equation}
r_S = \frac{1}{2} (1-h(e_X)-h(e_Z)),
\end{equation}
where $h$ is the binary Shannon entropy. The factor $1/2$ accounts for the use of two modes (in our case polarizations) per transmitted qubit. Note that this formula assumes perfect classical error correction and infinite length keys. Supporting the threshold introduced above, $r_S$ drops to $0$ at $e_X=e_Z=\unit[11]{\%}$. The resulting secret key rate is given in Fig.\ 3d, again for different cut-offs in the number of trials. While the overall rate is reduced by about two orders of magnitude due to the finite secret-key fraction and the factor four for the fraction of usable BSMs explained above, the scaling advantage still unfolds for increasing $n$.

This is further quantified in Fig.\ 3e where, similarly to the yield, the rate-versus-distance exponent is plotted for different cut-offs $n$. For small $n$, our protocol performs worse than direct transmission for which we assume no imperfections. This is due to the finite QBER which increases with distance and thus reduces the secret-key fraction for larger distances. However, with increasing $n$ the advantage given by the repeat-until-success strategy unfolds. Although the exponent does not reach the ideal limit as nicely as for the yield, our experiment beats the fundamental limitation regarding scalability of direct transmission. Further analysis of the results of Figure 3 can be found in \cite{Supplement}.

In summary, we have realized a quantum repeater node for unconditionally secure quantum key distribution and have observed a twofold improvement of the rate-versus-distance scaling. As an outlook, we address the question how much improvement is needed to beat direct transmission in absolute rate. Using the model described in Ref.\ \cite{Luong2016} we estimate that this is possible by doubling the memory time to $\unit[40]{ms}$, the BSM efficiency to 10\%, and increasing the protocol fidelity by 2\% \cite{Supplement}. For these parameters a repeater advantage unfolds for a communication length larger than $L\sim7L_\mathrm{att}$. At this distance the secret key rate then amounts to $5\times10^{-5}$ bits per channel use. The required improvements of the system performance seem feasible, especially for the memory time where values exceeding $\unit[100]{ms}$ have been achieved \cite{Koerber2018}. However, the average number of repeated trials has to be increased to about 150, a presently intolerably high value for which the atom would quickly be heated out of the trap. Hence, future experiments require better atom trapping and cooling, e.g., with optical tweezers \cite{Covey2019}. Once these improvements are implemented, we can investigate the scaling to a chain of repeater nodes \cite{Borregaard2015b,Uphoff2016}. Another perspective is to increase the number of atoms to more than two to boost the transmission rate \cite{Trenyi2020} or possibly link them to more than two communication partners, e.g., for achieving a quantum conference key agreement \cite{Murta2020}. In combination with quantum-logic gates \cite{Welte2018} for entanglement purification and heralded quantum memories \cite{Brekenfeld2020} as end nodes, the door towards full quantum repeaters seems open.

This work was supported by the Bundesministerium f\"{u}r Bildung und Forschung via the Verbund Q.Link.X (16KIS0870), by the Deutsche Forschungsgemeinschaft under Germany’s Excellence Strategy – EXC-2111 – 390814868, and by the European Union’s Horizon 2020 research and innovation programme via the project Quantum Internet Alliance (QIA, GA No. 820445).

%


\pagebreak

\onecolumngrid
\begin{center}
  \textbf{\large Supplementary Material: \\A Quantum Repeater Node Demonstrating Unconditionally Secure Key Distribution}\\[.2cm]
  S. Langenfeld, P. Thomas, O. Morin, and G. Rempe\\[.1cm]
  {\itshape ${}^1$Max-Planck-Institut f\"{u}r Quantenoptik, Hans-Kopfermann-Strasse 1, 85748 Garching, Germany\\}
(Dated: \today)\\[0.5cm]
\end{center}

\setcounter{equation}{0}
\setcounter{figure}{0}
\setcounter{table}{0}
\setcounter{page}{1}

\maketitle

\section{Methods}
The heart of the experiment are individual $^{87}$Rb atoms trapped in and strongly coupled to a high-finesse optical cavity with parameters $(g,\kappa,\gamma)/2\pi=\unit[(4.9,2.7,3.0)]{MHz}$ at wavelength $\lambda=\unit[780]{nm}$. Here, $\kappa$ and $\gamma$ are the cavity-field and atomic-dipole decay rates, respectively, and $g$ is the single-atom single-photon coupling constant at the center of the cavity and for both atomic transitions relevant in our protocol, i.e. $\ket{5^2S_{1/2},F=1,m_\mathrm{F}=\pm1}\leftrightarrow\ket{5^2P_{3/2},F'=1,m_\mathrm{F}=0}$ and $\ket{5^2S_{1/2},F=1,m_\mathrm{F}=0}\leftrightarrow\ket{5^2P_{3/2},F'=1,m_\mathrm{F}=\pm1}$. The cavity mirrors have different transmission coefficients ($\unit[100]{ppm}$ and $\unit[4]{ppm}$) so that photons predominantly leave the cavity via the higher-transmission mirror.

Atoms are transferred from a magneto-optical trap (MOT) into a two-dimensional (2D) standing-wave optical dipole trap, consisting of a red-detuned, $\lambda=\unit[1064]{nm}$ trap perpendicular to the cavity axis ($x$-axis) and a blue-detuned, $\lambda=\unit[772]{nm}$ trap along the cavity axis ($y$-axis). We take images of the atomic fluorescence via a high NA objective and an EMCCD camera. Starting from a random number of atoms at random positions, we use those images to select atomic patterns which are suitable for the experiment, namely two atoms at a suitable distance. If no such pattern exists, the loading process is restarted. Otherwise, all superfluous atoms are heated out of the trap by atom-selective application of a near-resonant beam. Thus, we end up with two atoms which have a suitable distance. The center of mass of these atoms can be moved along the $x-$axis by shifting the standing wave pattern. We place the atoms so that atom A is always at a fixed position, while the suitable distance is defined such that this results in both atoms being about symmetric to the cavity center. This guarantees maximum coherence times for atom A and about the same coupling strength of both atoms to the cavity.
The average trapping time of two atoms inside the cavity mostly depends on the amount of heating introduced by the repeated optical pumping for atom initialization. This depends on the number of necessary retrials to get a herald signal which is a function of the equivalent distance. The obtained trapping times for usable atomic patters, i.e. including image post-selection (which will be discussed below), for the three different loss scenarios depicted in the main text are $\unit[(8, 5, 4)]{s}$. 

Optical fields are either applied onto both atoms via a global beam ($w_\mathrm{R}\approx \unit[40]{\mu m}$) or atom-selectively via an optical addressing system ($w_\mathrm{R}=\unit[1.7]{\mu m}$) \cite{Langenfeld2020SM}. 
We thus chose a minimum inter-atomic distance of $\unit[8]{\mu m}$ to minimize cross-illumination of the addressing beam while still being close to the cavity center ($w_\mathrm{R} = \unit[29.6]{\mu m}$) in order to keep a high atom-cavity coupling.

\subsection{Detection setup for Alice, Bob and the Bell-state measurement}
In our proof-of-principle experiment, we use a single detection setup for Alice, Bob and the Bell-state measurement (BSM). In order to test the quantum repeater sequence in this configuration, fast polarization rotations are necessary so that the detection for Alice, Bob and the BSM can be performed in different polarization bases, i.e. the $X-$ or $Z-$basis. To this end, we use a DC-coupled, high-bandwidth electro-optical-modulator (EOM, Qubig PC3R-NIR) in front of the analyzing polarizing beamsplitter (PBS). The EOM allows to rotate the polarization in less than $\unit[1]{\mu s}$ and has a polarization extinction value of larger than $1:10000$, which guarantees the integrity of the polarization qubits.

A caveat of this configuration is that introducing transmission losses to Alice/Bob also affects the efficiency of the BSM. To avoid this, we use a software random number generator which for every individual photon generation attempt gates on or off the detectors of Alice and Bob conditioned on the outcome of a biased coin toss. The success probability of this coin toss can be adjusted to give the desired distance equivalent of the optical link. At the same time, the photon detection probability in the BSM is not affected. In this way, we effectively mimic the losses within the link to Alice and Bob while not affecting the BSM, which realistically models the underlying repeater protocol with the BSM performed in the immediate vicinity of the repeater node.

The actual losses of the detection setup, starting from the cavity output, can be summarized into an efficiency for the first fiber-coupling ($0.85$), optical elements of the detection setup ($0.75$) and the detectors themselves ($0.9$). For zero communication distance, $L=0$, this results in a total efficiency of $0.58$.

\subsection{Detailed repeater sequence}

\begin{figure}[b]\centering
\includegraphics{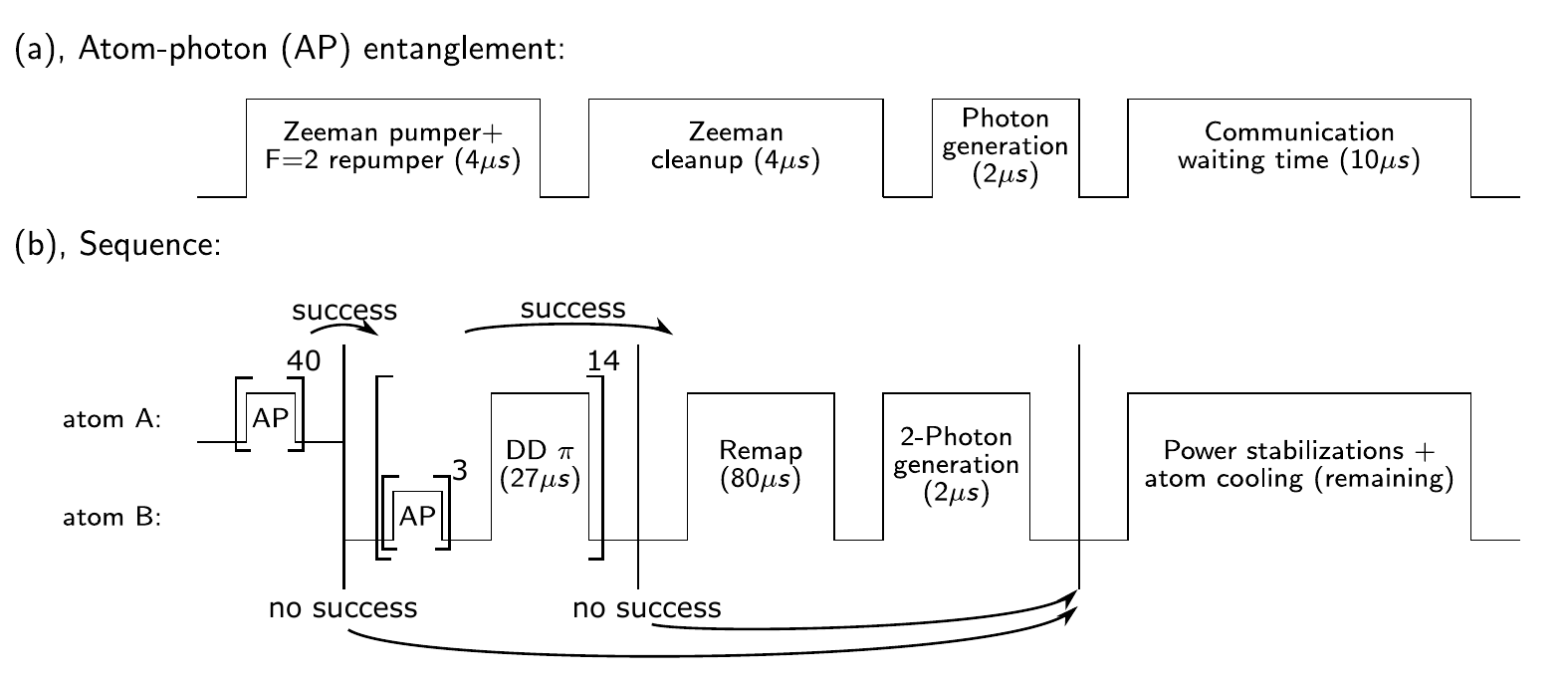}
\caption{Detailed experimental sequence.
(a) Description of the experimental steps to create atom-photon entanglement (AP) which is used for establishment of correlation between atom A/B and Alice/Bob, respectively. 
(b) The combined quantum repeater sequence, including the entanglement sub-sequence from part (a). Square brackets denote the repetition of this sub-sequence until the transmission was successful. The exponent gives the maximum number of repetitions. If this number is reached (no success), the repeater protocol is aborted and the sequence proceeds to utility tasks until the next run is started. If the transmission was successful (success), the protocol proceeds to the next stage.
}
\label{fig:seq}
\end{figure}

Each attempt to generate atom-photon (AP) entanglement starts with atom initialization to the ground state $\ket{F=2, m_\mathrm{F}=0}$. The pumping to this specific Zeeman state is achieved with a $\pi-$polarized beam which is close to resonance to the transition $\ket{5^2S_{1/2}, F=2, m_\mathrm{F}} \leftrightarrow \ket{5^2P_{1/2}, F'=2, m_\mathrm{F}}$. As this transition is dipole-forbidden for $m_\mathrm{F}=0$, population accumulates in the target initial state. At the same time, population in $\ket{5^2S_{1/2}, F=1}$ has to be repumped to $\ket{5^2S_{1/2}, F=2}$ which is accomplished with another $\pi-$polarized beam close to resonance to the transition $\ket{5^2S_{1/2}, F=1, m_\mathrm{F}} \leftrightarrow \ket{5^2P_{3/2}, F'=2, m_\mathrm{F}}$. This combination of beams is applied for $\unit[4]{\mu s}$ (Fig. \ref{fig:seq}a). In the atom-photon entanglement scheme, every state in the $F=2$ manifold can generate photons. In order to remove potentially remaining population from all unwanted states, the same Zeeman pumper is applied again for $\unit[4]{\mu s}$ without the additional repumper. This depopulates all states in the $F=2$ manifold except of the one with $m_\mathrm{F}=0$, leading to a high-fidelity state preparation.
After initialization, the photon-generating vacuum-stimulated Raman adiabatic passage (vSTIRAP) pulse is applied. The pulse power is tuned such that the emitted photon has a temporal full width at half maximum (FWHM) of about $\unit[300]{ns}$. In order to simulate the transmission time of this photon as well as the classical communication time between Alice/Bob and the repeater node, the sequence pauses for $\unit[10]{\mu s}$ which equals a maximum distance between repeater node and Alice/Bob of $\unit[0.9]{L_\mathrm{att}}$ for the wavelength $\lambda=\unit[780]{nm}$ using $L_\mathrm{att}(\unit[780]{nm}) = \unit[1.1]{km}$.

The whole sequence now incorporates the above elements (Fig. \ref{fig:seq}b). First, up to $40$ atom-photon entanglement attempts are repeated on atom A. If no herald signal arrived from Alice within those $40$ trials, the sequence jumps to atom cooling and optical beam power stabilizations.
If Alice heralds a photon detection, the protocol jumps to the next part of the sequence. Here, atom-photon entanglement attempts for atom B are combined with dynamical decoupling (DD) pulses for atom A. As the DD pulses are not atom-selective, they have to be interleaved with the attempts on atom B. We use a combination of three atom-photon entanglement attempts before one dynamical decoupling pulse is applied. This combination is repeated up to $14$ times, giving a total of $42$ attempts on atom B. Again, if none of the attempts succeeded, the sequence jumps to atom cooling and power stabilizations. However, as soon as one of the photons arrived, the sequence jumps to the BSM, which consists of a remapping pulse to transfer the population to $\ket{F=2}$ again and subsequent photon generation. This completes the attempt to generate one bit of a shared key. Afterwards, the sequence regularly proceeds to atom cooling and power stabilizations. The whole protocol is repeated at a rate of $\unit[160]{Hz}$.

\newpage
\section{Coherence time and dynamical decoupling}

\begin{figure}[b]\centering
\includegraphics{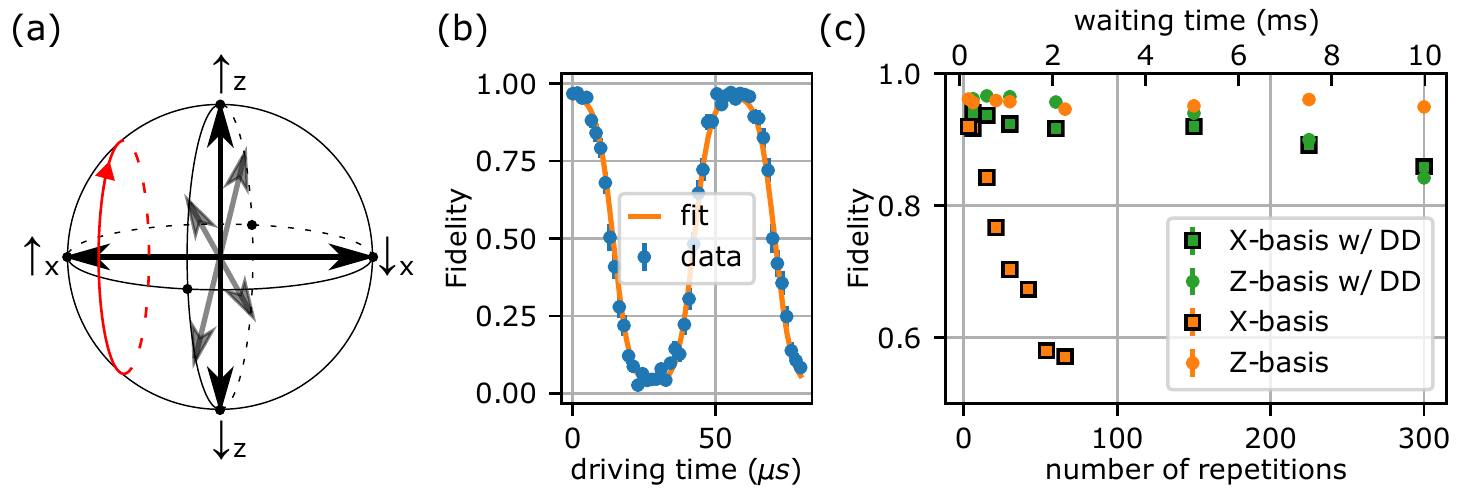}
\caption{Dynamical decoupling for extending coherence time.
(a) Bloch sphere of the atomic qubit and the rotation due to the dynamical decoupling pulses around the $X-$axis. Thus, qubits along the $Z-$axis experience a bit-flip while qubits along the $X-$axis are not modified. 
(b) Rabi flopping for states prepared along the $Z-$axis. Due to the three-level nature of the driving process (see Fig. 2b of the main text), the data and fit do not follow a simple sinusoidal curve. 
(c) Experimental results for the achievable coherence time with (green) and without (orange) the decoupling sequence employed for the experiments presented in the main text, namely a $\pi$-pulse around the $X-$axis every $\unit[99]{\mu s}$. Here, number of repetitions refers to the number of atom-photon entanglement attempts in the main text. We apply a DD pulse every three attempts (see Methods section).
}
\label{fig:DD}
\end{figure}

In the protocol, atom B repeatedly tries to connect to Bob, while the previously established correlation between Alice and atom A needs to be preserved. Due to the requirement of keeping the quantum bit error rate (QBER) below 11\%, the coherence time of atom A needs to be much longer than the time it takes to establish a correlation between atom B and Bob. Depending on photon generation and detection efficiencies as well as photon propagation losses, the coherence time needs to be orders of magnitude longer than the classical communication time between the repeater node and Bob.

Thus, for extending the coherence time of the qubit memory utilized in the repeater protocol, we employ dynamical decoupling (DD). 
Due to the special setting of BB84, only four different states need to be decoupled from environmental noise. Thus, for reasons of simplicity, we employ a CP/CPMG sequence \cite{CP1954SM,MG1958SM}, in which rotations are only performed along a given axis which is either perpendicular or parallel to the qubit state. As $(\ket{\uparrow_Z},\ket{\downarrow_Z})=(\ket{F=1,m_\mathrm{F}=1},\ket{F=1,m_\mathrm{F}=-1})$ are energy eigenstates of $^{87}$Rb, we perform the DD rotations along the $X-$axis (Fig. \ref{fig:DD}a).
To apply these pulses, we use a double two-photon Raman transition whose single-photon detuning is set to a point at which off-resonant scattering is minimized ($\lambda \approx \unit[790]{nm}$). As such a far-detuned Raman configuration interferes destructively for $\Delta m_\mathrm{F}=2$ transitions, the two-photon detuning is set to a single Zeeman splitting ($\delta_L/2\pi = \unit[50]{kHz}$) of the $F=1$ ground-state. Thus, transitions between $\ket{F=1,m_\mathrm{F}=\pm1}$ are driven via the intermediate state $\ket{F=1,m_\mathrm{F}=0}$ (see also Fig. 2b of the main text).
In order to not drive transitions which are detuned by two Zeeman splittings from the intended transition, i.e. $\Delta m_\mathrm{F}=-1$, the driving rate has to be smaller than the Zeeman splitting. As is depicted in Fig. \ref{fig:DD}b, the Rabi frequency is $2\pi \times \unit[17.6]{kHz}$.

Experimental results are shown in Fig. \ref{fig:DD}c. Without dynamical decoupling, the coherence time is limited to about $\unit[1]{ms}$, mostly given by magnetic field fluctuations and varying AC Stark shifts given by the optical dipole trap. Dynamical decoupling can effectively compensate these fluctuations which leads to an estimated coherence time of more than $\unit[20]{ms}$. Note that in contrast to the case without decoupling, also the energy eigenstates ($Z-$basis) decohere with increasing number of DD pulses. This can be explained by residual off-resonant light scattering from the Raman laser. The limitation to 300 pulses is exerted by our field programmable gate array (FPGA) control hardware of the experiment.

Previous achievements on the same setup used decoherence-free substates \cite{Koerber2018SM} to achieve a coherence time exceeding $\unit[100]{ms}$. However, this came at the cost of cooling to the motional ground-state, less flexibility in the readout time of the qubits and a reduced starting fidelity at $t=0$. As preparation time and total fidelity are very important parameters for the quantum repeater protocol, we chose dynamical decoupling instead. Note that the obtained coherence time is not the primary limitation of the achievable rate and distance, as we currently only use up to $40$ repetitions.

In summary, we found that we could minimize the increase in QBER from 10\% without dynamical decoupling to below 1\% with dynamical decoupling during the up to $40$ repetitions implemented in our experiment. Thus, dynamical decoupling is one of the enabling technologies for achieving unconditional security in our experiment.

\section{Bell-state measurement}

\begin{figure}[t]\centering
\includegraphics{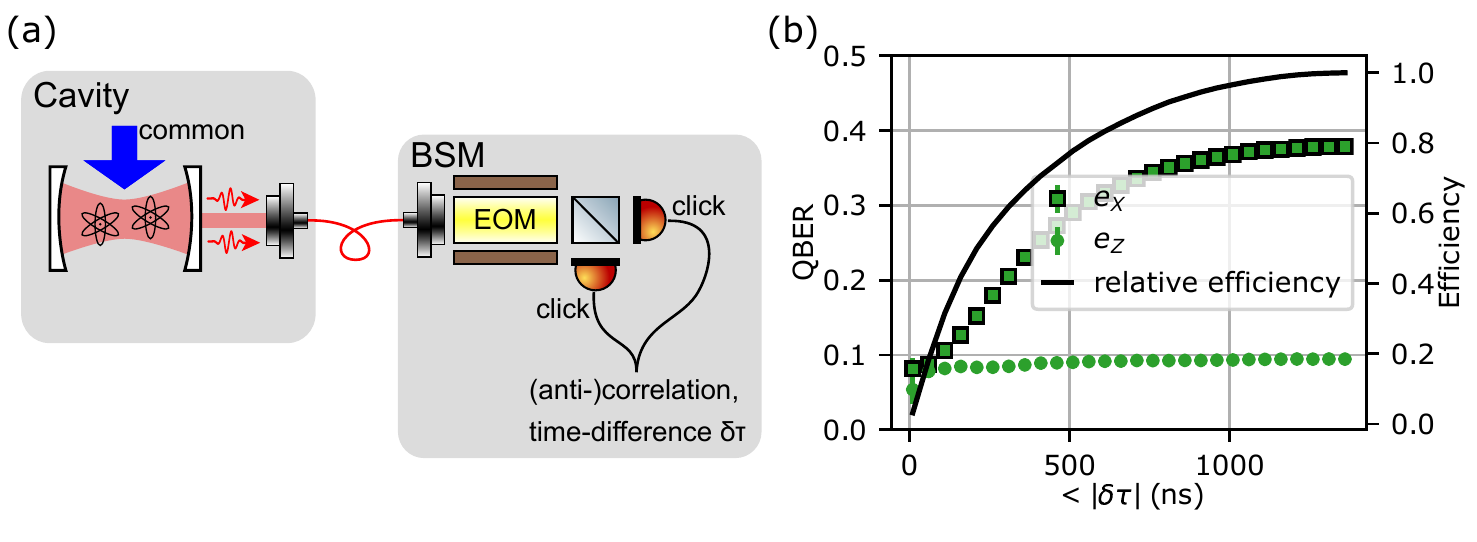}
\caption{BSM: evolution of QBER. 
(a) Sketch of the detection of two photons from two atoms, with the cavity erasing the which-way information. The detection of two orthogonal polarized photons after a polarizing beamsplitter heralds the projection onto $(\ket{\uparrow_Z \downarrow_Z} + \ket{\downarrow_Z\uparrow_Z})/\sqrt{2}$. This leads to (anti-) correlation for Alice and Bob if they measured in the ($Z-$) $X-$basis. 
(b) Quantum bit error rate (QBER) for detection clicks within a maximum time difference $\delta \tau$. Results for Alice and Bob measuring in the $X$-basis (squares) and in the $Z$-basis (circles) are shown as well as the relative efficiency of the BSM (line).
}
\label{fig:BSM}
\end{figure}

After Alice and Bob have measured their photons in a randomly chosen basis, the two atoms are in the product state $(\alpha_1 \ket{\uparrow_Z} + \beta_1 \ket{\downarrow_Z}) \otimes (\alpha_2 \ket{\uparrow_Z} + \beta_2 \ket{\downarrow_Z})/2$ where the indicies enumerate the two atoms.
The photonic readout process maps these states onto photons (Fig. \ref{fig:BSM}a). Due to the bosonic nature of the ideally indistinguishable photons, only $\ket{\Phi^+}, \ket{\Phi^-}, \ket{\Psi^+}$ can enter the same spatial mode of the cavity. The detection of two different polarizations finally heralds the projection of the shared Alice-Bob state onto $\ket{\Psi^+}$. However, if the two photons are not perfectly indistinguishable, or if a phase evolution of the atoms in-between the two photon emissions occurs due to some process, the detection of $\ket{\Psi^-}$ states becomes possible. The potential correlation outcomes for Alice and Bob for these two Bell states are given in Table 1. 

\begin{table}[hbt]\centering
\caption{Potential correlation outcomes for Alice and Bob if the BSM heralded one of the two Bell states $\ket{\Psi^\pm}$.}
	\begin{tabular}{| c || c | c | c | c || c | c | c | c |}
	\hline
	 & $\uparrow_X \uparrow_X$ & $\uparrow_X \downarrow_X$ & $\downarrow_X \uparrow_X$ & $\downarrow_X \downarrow_X$ & $\uparrow_Z \uparrow_Z$  & $\uparrow_Z \downarrow_Z$  & $\downarrow_Z \uparrow_Z$  & $\downarrow_Z \downarrow_Z$ \\
	\hline
	$\ket{\Psi^+}$ & 0.5 & 0 & 0 & 0.5 & 0 & 0.5 & 0.5 & 0\\
	$\ket{\Psi^-}$ & 0 & 0.5 & 0.5 & 0 & 0 & 0.5 & 0.5 & 0\\
	\hline
	\end{tabular}
\end{table}

If Alice and Bob both measured in the $Z-$basis, the distinction between $\ket{\Psi^\pm}$ does not matter as they both lead to anti-correlation for Alice and Bob. However, if they measured in the $X-$basis, $\ket{\Psi^+}$ leads to correlation while $\ket{\Psi^-}$ leads to anti-correlation for Alice and Bob. The consequence of this can be seen in Fig. \ref{fig:BSM}b. For small maximum time differences between the two BSM clicks, the photons are nicely indistinguishable and no phase-evolution has occurred. Thus, the quantum bit error rate is small for both detection bases. However, with increasing $\delta \tau$, the detection of $\ket{\Psi^-}$ becomes possible which increases the QBER for the detection in the $X-$basis while it stays about constant for the detection in the $Z-$basis. As Alice and Bob have full information about these aspects in their classical post-processing phase of the quantum key distribution protocol, they can choose a trade-off between minimum QBER and decreasing relative efficiency which maximizes the secret key rate, which is also what we do in the results presented in the main part of this work.

\section{Data processing}
\begin{figure}[t]\centering
\includegraphics{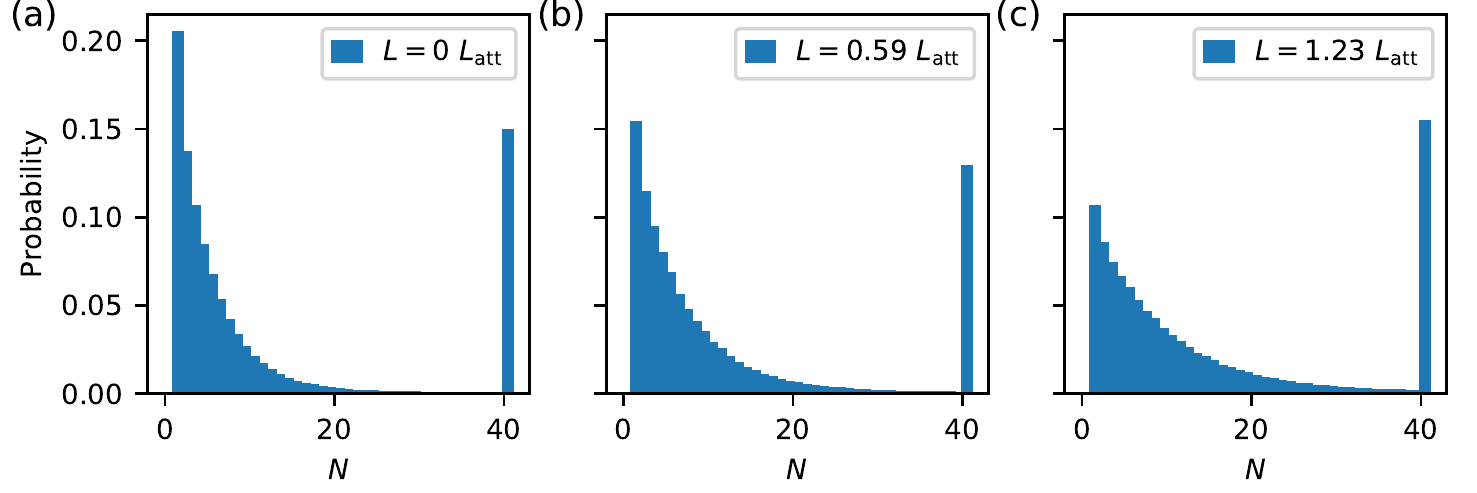}
\caption{Number of trials necessary to establish correlation. 
(a, b, c) Distribution of the necessary number of trials until Bob heralds the detection of a photon for the three different communication distances indicated in the figure.
}
\label{fig:trial_distro}
\end{figure}

Here we outline a number of steps that we took to process the acquired data.
\paragraph{Loss of atoms}
During an experimental run, we acquire an image of the atoms every $\unit[500]{ms}$ to check for their positions. This time is necessary to acquire enough fluorescence light from the atoms in order to have a sufficiently good signal-to-noise ratio on the camera.
Due to the limited trapping times of the atoms, the last image is likely to have a certain duration in which one of the atoms has disappeared but is still visible on the image. During this time, the transmission to Alice or Bob will never be successful as there is no atom which could emit a photon. This becomes apparent when looking at the histogram of the number of necessary atom-photon entanglement attempts until a heralding signal is received (Fig. \ref{fig:trial_distro}): 
The figure shows a contribution at the maximum number of trials (40) although the regular exponential has decayed long before this. As this effect is only a by-product of the limited trapping time, we only use events with $N <40$ which modifies the rate per channel use by an observed factor of $0.83$. This factor is more or less distance-independent and thus does not modify the conclusions drawn on the rate-distance exponent.

\paragraph{Equivalent distance}
As is described in the methods section above, the transmission distance for different data points in Fig. 3 of the main text is simulated by introducing artificial losses. These are reliable and do not change over time. However, additional losses in the setup might occur due to the mechanical drift of fiber couplings in the optical pathway to Alice or Bob. These drifts slightly change the losses which are relevant for the equivalent distance. We thus calculate equivalent distance by using the known relation $p_\mathrm{AB} = 1/\langle N \rangle$ where $p_\mathrm{AB}$ is the single-trial success probability and $\langle N \rangle$ is the average number of trials necessary to have a successful transmission to Alice or Bob. Then, the equivalent distance is given by $L/L_\mathrm{att}=-2\ln(p_\mathrm{AB})=2\ln(\langle N \rangle)$. As $\langle N \rangle$ can be reliably extracted form the accumulated data, we use this quantity to calculate the equivalent distance. However, this number includes the losses of the experimental setup which are not accountable to distance. As we know that the first data point corresponds to zero additional losses, we subtract a constant positive offset along the distance axis such that the equivalent distance of the first data point is zero on the plot. In our view, this represents the most conservative and realistic scenario, where only losses introduced beyond the experimental imperfections are considered as equivalent distance.

\paragraph{BSM efficiency}
The data for different distances (Fig. 3 of the main text) is taken over the course of multiple weeks. During this time, the efficiency of the BSM drifted by about $\unit[10]{\%}$ due to mechanical drifts of multiple fiber couplers in the optical pathway. It turns out that this resulted in a higher efficiency for the data points of larger distance, thus seemingly magnifying the effect of the repeater advantage. Without the compensation described below, the points given in Fig. 3b and 3e of the main text even surpass the repeater bound of $-0.5$. 
For the compensation, we evaluate the BSM efficiency for each distance individually ($p_\mathrm{BSM,L}$). As the key rate scales linearly with a given BSM efficiency, we then multiply the obtained rates given in the main text with $\langle p_\mathrm{BSM} \rangle/p_\mathrm{BSM,L}$, where $\langle p_\mathrm{BSM} \rangle=\unit[(5.07\pm0.03)]{\%}$ is the average of all $p_\mathrm{BSM,L}$. 
With this renormalization we arrive at a scenario in which the BSM efficiency is constant over time.

\paragraph{Additional key rate contributions}
In the main text of this work, the yield and secret key rate are given as the number of obtained bits per channel use, assuming that the atoms are ready for the repeater protocol.
Here we will give additional experimental parameters which relate to the data rate.
The first contribution is already given in the above paragraph describing the loss of atoms, namely the selection on $N<40$ which modifies the rate per channel use by an observed factor of $0.83$. 
Another contribution is due to the atom positioning within the cavity. Although we do perform atom pre-selection based on the fluorescence images, we also perform post-selection on the images to ensure having the right distance and absolute positions of the atoms. On average, $\unit[42]{\%}$ of the images survive this selection which directly enters the duty cycle for key generation.
Last, new atoms have to be reloaded or the current atoms have to be repositioned whenever they do not match the given criteria. This results in a duty cycle of $\unit[(25,20,15)]{\%}$ for the three different distances given in the main text.
In total, the product of the three factors results in an average duty cycle of $c_\mathrm{dc} \approx 0.83 \times 0.42 \times 0.2\approx 0.07$.
The real-time yield, i.e. the rate of not yet secured bits per time, then evaluates to:
\begin{equation}
    p_\mathrm{BSM} \times \Gamma_\mathrm{seq} \times c_\mathrm{dc} \approx 0.05\frac{\mathrm{bits}}{\mathrm{attempt}} \times 160\frac{\mathrm{attempts}}{s} \times 0.07 \approx 0.57\frac{\mathrm{bits}}{s}.
\end{equation}

\paragraph{Error analysis for the data presented in Fig. 3 of the main text}
The yield (Y) given in Fig. 3a represents a Bernoulli trial with $k\approx10^6$ trials per data point, with the exact number depending on $n$.
The error bars are then given by $\sqrt{Y(1-Y)/k}$.
The same holds for the estimation of the quantum bit error rates depicted in Fig. 3c. Exemplary described for $e_Z$, the QBER is given by $e_Z=k_\mathrm{corr}/(k_\mathrm{corr}+k_\mathrm{acorr})$, where $k_\mathrm{corr}$ ($k_\mathrm{acorr}$) is the number of bits showing (anti-)correlation for Alice and Bob given they both measured in the $Z-$basis. The total number of bits is then $k_\mathrm{total}=k_\mathrm{corr}+k_\mathrm{acorr}$.
Thus, in this Bernoulli trial the error bars are given by $\sqrt{e_Z(1-e_Z)/k_\mathrm{total}}$. The total number of bits per data point is typically on the order of $k_\mathrm{total}\approx 1000$.
The secret key rate depicted in Fig. 3d is the product of the secret key fraction (Eq. 1 in main text) and the yield, taking into account the reduction of the yield due to the limited usability of the BSM as described in the main text and the supplementary text on the BSM.
The error of the secret key fraction as a function of $e_X$ and $e_Z$ is calculated according to error propagation assuming independent variables. 
Note that the correlation between errors in the QBER and the resulting secret key fraction is about 10, i.e. $\delta r_S/\delta e_{X,Z}\approx 10$. Although all obtained QBERs beat the threshold of $\unit[11]{\%}$ by more than three standard deviations, the error bar of the secret key fraction is still quite large.
For the error bars of the secret key rate, we use error propagation assuming independent variables. The larger error bar of the secret key fraction dominates the error of the secret key rate and leads to the errors depicted in Fig. 3d.
Figures 3b and 3e show the slope of linear fits to the data presented in Figs. 3a and 3d, respectively. In order to model the slope and its error faithfully, we use Monte Carlo simulations assuming the data points (error bars) of Figs. 3b and 3e represent the mean (one standard deviation) of a normally distributed random variable. We use linear fits to $1000$ samples drawn from these distributions to obtain distributions for the slopes. From these distributions we infer the mean and standard deviation which we finally depict in Figs. 3b and 3e.

\section{Outlook}

\begin{figure}[b]\centering
\includegraphics{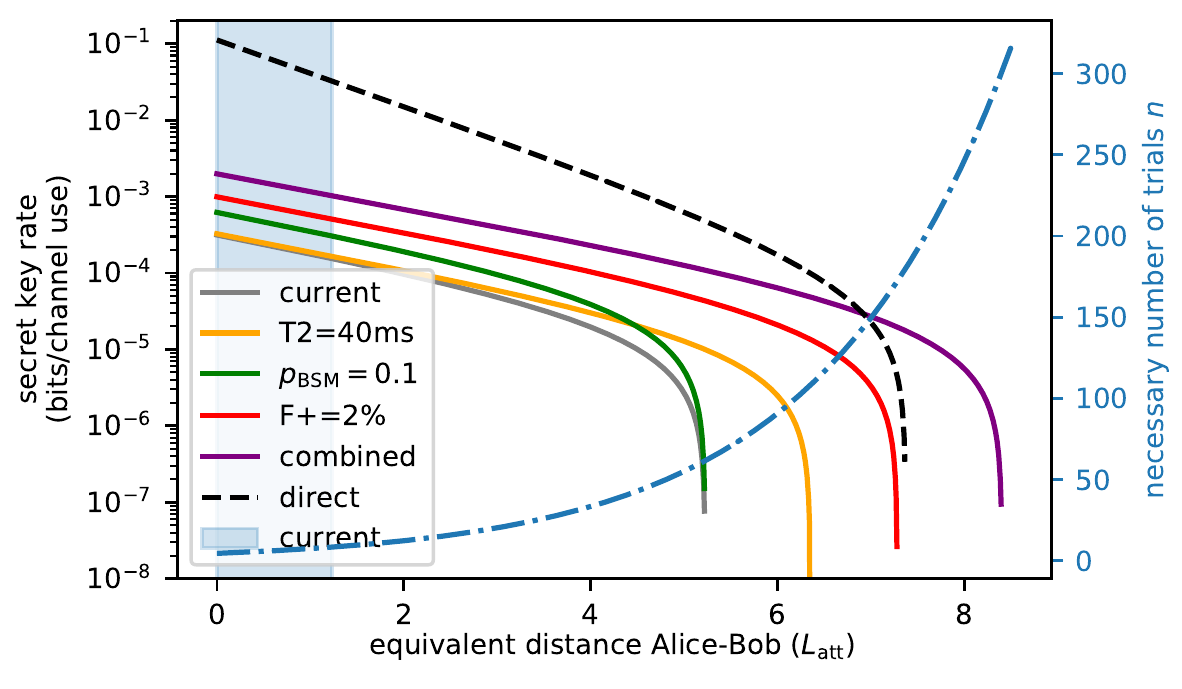}

\caption{Outlook on secret key rate achievable with future improvements to the setup. The blue-shaded region shows the range of investigated equivalent distance, limited by the necessary average number of retrials $n$ (right vertical axis, dash-dotted line). Except of the purple one, the individual curves show the achievable secret key rate if only the highlighted parameter is changed compared to the current setup. The purple curve shows the result of the combination of the individual improvements. All curves are calculated by using the formulas given in Ref.\ \cite{Luong2016SM} and the numbers summarized in this section.
}
\label{fig:outlook}
\end{figure}

In the main manuscript, we give an outlook on which improvements are necessary to beat direct transmission in an absolute-rate scenario. This outlook is based on calculations using the formulas given in Ref.\ \cite{Luong2016SM}. As input parameters we use the currently achieved parameters as presented in this manuscript and summarized in the following. For easier reference, we also include the labeling of Ref.\ \cite{Luong2016SM}:
\begin{itemize}
    \item $T_\mathrm{prep}=\unit[10]{\mu s}$ (preparation time)
    \item $\eta_\mathrm{tot} = \eta_c \times \eta_d \times \eta_\mathrm{prep} = p_\mathrm{AB,L=0}=\unit[22.13]{\%}$ (zero-distance total efficiency)
    \item $T_2 = \unit[20]{ms}$ (dephasing time)
    \item $e_\mathrm{mA,mB} = 0$ (misalignment error)
    \item $p_d = \unit[20]{Hz}\times \unit[0.5]{\mu s}=10^{-5}$ (dark-count probability per detector)
    \item $p_\mathrm{BSM}=\unit[5.07]{\%}$ (BSM efficiency)
    \item $f=1$ (perfect error correction)
    \item $F=0.925$ (zero-distance fidelity).
\end{itemize}

Figure \ref{fig:outlook} shows the expected secret key rate versus equivalent distance for the current scenario, for scenarios in which individual parameters are improved as well as a combination of those improvements. The direct-transmission rate drops for large distances when the dark-count probability becomes comparable to the photon click rate, so that the QBER increases also when using direct transmission. As stated in the main text, the repeater beats direct transmission if the coherence time and the BSM efficiency can be doubled, while the fidelity of the overall protocol has to be improved by 2\%. In that case, the repeater can maintain its advantageous scaling for a sufficiently long distance. Note that the repeater is less prone to dark counts, as the individual communication distances between the repeater node and Alice/Bob are only half the distance compared to direct transmission. Thus, the crossing point for beating direct transmission increases to larger distances when reducing the dark count probability, resulting in even higher demands on the repeater hardware in order to beat direct transmission.


\end{document}